\documentclass[iop]{emulateapj}

\shorttitle{Heating Extrasolar Giant Planets}
\shortauthors{Buzasi}

\begin{document}


\title{Stellar Magnetic Fields as a Heating Source for Extrasolar Giant Planets}


\author{D. Buzasi}
\affil{College of Arts and Sciences, Florida Gulf Coast University,
    Fort Myers, FL 33965}
\email{dbuzasi@fgcu.edu}


\begin{abstract}
It has been observed that hot Jupiters located within $0.08 \rm~AU$ of their host stars commonly display radii in excess of those expected based on models. A number of theoretical explanations for this phenomenon have been suggested, but the ability of any one mechanism to account for the full range of observations remains to be rigorously proven. I identify an additional heating mechanism, arising from the interaction of the interplanetary magnetic field and the planetary magnetosphere, and show that this is capable of providing enough energy to explain the observed planetary radii. Such a model predicts that the degree of heating should be dependent on the stellar magnetic field, for which stellar activity serves as a proxy. Accordingly, I examine populations of hot Jupiters from the {\it Kepler} database and confirm that stellar activity (determined using {\it Kepler} CDPP levels) is correlated with the presence of planetary radii inflated beyond the basal level of $R = 0.87 R_J$ identified by previous researchers. I propose that the primary mechanism for transferring energy from the magnetosphere to the planetary interior is Joule heating arising from global electric circuits analogous to those seen in solar system objects.
\end{abstract}


\keywords{stars: planetary systems --- planets and satellites: general --- planets
and satellites: atmospheres}



\section{Introduction}

Giant planets are intrinsically interesting both because they enable us to better understand the formation and evolution of our own solar system and because they serve as laboratories for the science of materials at temperatures and pressures beyond those achievable in the laboratory (Perna, Heng, \& Pont 2012). The study of extrasolar giant planets became an observational discipline in 1995, with the discovery of the first example, 51 Peg b (Mayor \& Queloz 1995). Four years later, Charbonneau {\it et al.} (2000) found the first transiting extrasolar giant planet, HD 209458b. The existence of transits enabled measurement of the planetary radius, which Charbonneau et al. found to be inflated beyond both expectations based on models and those based on our experiences in our own solar system. Since that time, approximately 200 hot Jupiters have been discovered, and inflated radii have proven to be a common, although not universal, phenomenon (Demory \& Seager 2011).

Planets in this size range are supported by a combination of electron degeneracy and thermal pressure, so the resulting models predict an approximately constant radius for planets with masses between $\sim 1$ and $\sim 7 \rm M_{J}$. Quite early on, Guillot et al. (1996) noted that, while 
giant planets far from their host stars tend to follow the expected mass-radius relation, those closer in are typically inflated in comparison to model predictions. The most recent study in this area is by Demory \& Seager (2011), who examined a {\it Kepler} sample of 138 giant planet candidates, and determined that those experiencing incident stellar fluxes less than $\sim 2 \times 10^8 \rm~erg~cm^{-2}~s^{-1}$ showed constant radii of $0.87 \pm 0.12 \rm R_J$, while planets experiencing stellar fluxes above this level were inflated, with the degree of inflation increasing along with the stellar flux level.

Several models have been put forward to explain these observations. A more detailed explication is given in Fortney \& Nettlemann (2010), but essentially most models require either modifications to improve heat retention, or additional energy sources. Thus, Burrows {\it et al.} (2007) suggest enhanced atmospheric opacities, while Chabrier \& Baraffe (2007) show that heavy element abundance gradients could suppress convection, both leading to greater radii. Alternatively, Hansen \& Barman (2007) suggest that preferential He mass loss would lead to H-rich compositions, and thus greater radii. In the second category, additional energy sources have been posited to arise from tidal heating (Bodenheimer {\it et al.} 2001; Gu, Bodenheimer, \& Lin 2004), enhanced conversion of the incident stellar flux to planetary wind energy (Guillot \& Showman 2002), or interaction of ionized planetary winds with the planetary magnetic field (Batygin \& Stevenson 2010). The amount of additional heating required can be as large as $8 \times 10^{27} \rm~erg~s^{-1}$ (for Tres-4b; see Batygin \& Stevenson 2010), and the additional heating must be supplied to the deep interior of the planet, typically below the $10^2 - 10^3 \rm~bar$ level. Furthermore, some proposed models have difficulties in accounting for the increase in radii for large incident stellar fluxes (Laughlin {\it et al.} 2011), although Batygin {\it et al.} (2011) show that the ``Ohmic dissipation'' model, whereby heating derives from interactions between atmospheric flows and the planetary magnetic field, can successfully reproduce this phenomenon.

The critical incident stellar flux level of $\sim 2 \times 10^8 \rm~erg~cm^{-2}~s^{-1}$ identified in Demory \& Seager translates, in the case of a star like the Sun, to a circular orbit with a semimajor axis of approximately 0.08 AU, or $18 \rm R_\sun$. It is suggestive that this roughly corresponds to the Alfv\'{e}n radius for the Sun, variously estimated at $5 - 50 \rm R_\sun$ (Aib\'{e}o {\it et al.} 2007, Scrijver, DeRosa, \& Title 2003). In our solar system, magnetospheric extraplanetary energy input at both the Earth and Jupiter is dominated by kinetic energy from the solar wind, but in this paper I suggest that for the much closer hot Jupiters the magnetic energy component of the stellar wind is a more important contributor and may be capable of providing enough energy to account for the inflated radii which are observed for hot Jupiters close to their host stars.

\section{Model}


Any planet with an internal magnetic field can be expected to be surrounded by a magnetosphere, which serves as an energy reservoir, with energy flowing in from the stellar wind and the planetary rotation, and out via heating of the solar wind and planet through a number of mechanisms (Vasyliunas 2010). In this work, I neglect the influence of planetary rotation, and consider a planet with radius $\rm R_{PL} = R_J$ and a magnetic dipole field of $1-25 \rm G$.

Pressure balance considerations then yield the scale size of the magnetosphere. Stellar wind pressure has contributions from the stellar magnetic pressure $P_B = B^2/{8 \pi}$, the kinetic (ram) pressure $P_k = \rho {v_w}^2 / 2$, and thermal pressure $P_{th} = n k T$; the latter can be neglected as it is at least two orders of magnitude less than the other contributions over the entire range of interest. In the case of both the Earth and Jupiter, the ram pressure term dominates over stellar magnetic pressure, but the presence of multipole components in the stellar magnetic field implies that this may not be the case for small orbital radii. Balancing stellar wind pressure against the pressure from the planetary dipole magnetic field gives
\begin{equation}
R_M = R_{PL} \left\{ \frac{B_{PL,0}^2}{{B_w}^2 + 4 \pi \rho {v_w}^2} \right\}^{1/6}
\end{equation}
This is a generalization of the Chapman-Ferraro radius (Chapman \& Ferraro 1931), such that the pressure of the stellar wind is applied over the area $\sigma = \pi {R_M}^2$.

I model the stellar wind using the model outlined in Suzuki (2006), with the magnetic field given by the analytic description in Banaszkiewicz, Axford, \& McKenzie (1998). This relatively simple model gives results similar to those seen in more complex implementations for radii outside the source surface (taken to be $\rm R_{SS} = 2.5 R_*$), and in any case results here are insensitive to details of the model used. The calculated field $(B_\rho, B_z)$ is presumed to possess cylindrical symmetry; in addition, I limit myself to the magnetic field and wind in the equatorial plane of the star (presumed to be the orbital plane of the planet), but the model is easily extended to more complex (and realistic) cases. I followed Banaszkiewicz, Axford, \& McKenzie in adopting $K = 1.0$, $M = 1.789$, $a_1 = 1.538$, and $Q = 1.5$ in order to force the wind model into close correspondence with the observed solar wind.

I adopt a commonly-used empirical formula describing the rate of terrestrial magnetospheric heating by storm and sub-storm processes (the so-called ``epsilon parameter'', see, {\it e.g.} Stern 1984):
\begin{equation}
\epsilon = v {l_0}^2 B^2 \sin^4 \theta
\end{equation}
Here $\theta$ is the polar angle of the solar interplanetary magnetic field vector, projected onto the Y-Z plane (Akasofu 1981). Such a model can be defended on dimensional grounds, as well as in more detailed arguments (Vasyliunas 2010), which identify the parameter as the power generated by the dynamo defined by the interaction of solar wind and magnetosphere. Note that $l_0$ is a scale length for the magnetosphere. Early work on the terrestrial magnetosphere adopted $l_0 = 7R_\earth$, but more recent analysis (Tanskanen et al. 2005) has adopted a revised scaling of $l_0 = 10R_\earth$. Noting that (unsurprisingly) $l_0 \approx R_M$ for both the Earth and Jupiter, in this work I adopt $l_0 = R_M$.

\section{Results}

Figure 1 illustrates the rising importance of the magnetic energy component of the stellar wind pressure for smaller semi-major axes. For orbital distances below about 0.15 AU, it is dominant. Note in particular that at $a = 0.08 \rm AU$, the available wind magnetic power is approaching $10^{27} \rm~erg~s^{-1}$ and rising rapidly; this is just the point at which Demory \& Seager (2011) see hot Jupiter radii start to increase. In addition, the power supplied by the wind is of the correct order of magnitude to supply the ``missing'' planetary heating component.

\begin{figure}
\epsscale{1.0}
\plotone{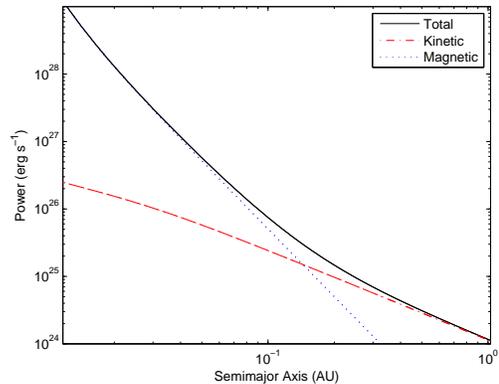}
\caption{Contributions to the total stellar wind energy available to the planetary magnetosphere. The kinetic energy component is shown in red (dot-dash), while the magnetic contribution is in blue (dotted); total wind energy is given in black (solid). The thermal energy component is not shown as it is more than two orders of magnitude smaller than the others over the range of the figure. Note that for semimajor axes below about $30 R_{\sun}$, corresponding to approximately $0.15~\rm AU$, the magnetic contribution dominates. In all figures, the abscissa runs from $a = R_{SS}$ to $a = 1\rm~AU$.}
\end{figure}

Of course, this agreement could simply be a fortuitous consequence of well-chosen values for the two main free parameters, $B_{PL}$ and $B_*$. Decreasing the planetary magnetic dipole strength $B_{PL}$ leads to smaller planetary magnetospheres, and thus to less magnetospheric energy deposition. However, since as $a \rightarrow R_{SS}$, $R_M \propto B_{PL}^{1/6}$, the effect is quite small, and thus the energy deposition seen in the model is relatively insensitive to $B_{PL}$.

The effects of changing the stellar magnetic field are much more substantial, as shown in Figure 2, which illustrates the effects of increasing the stellar magnetic field by factors of 10 and 100, while maintaining all other parameters constant. Cranmer \& Saar (2011) suggest that the photospheric magnetic filling factor is related to the stellar Rossby number by $f_* \sim \rm Ro^{-n}$, where $\rm n$ lies between 2.5 and 3.4. If the magnetic field at $R_{SS}$ scales similarly, this would imply rotation rates of $4 - 6$ days and filling factors near saturation for the most extreme example shown. The magnetic component of wind power which is available to the magnetosphere in this case can be as large as $10^{30} \rm~erg~s^{-1}$ at $a = 0.04 \rm AU$.

\begin{figure}
\epsscale{1.0}
\plotone{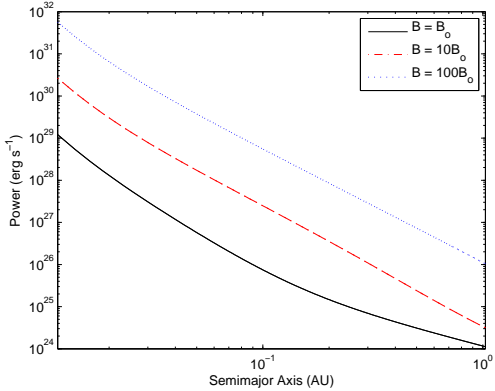}
\caption{The effects of changing the stellar magnetic field. The solid (black) line shows the solar case, as in Figure 1, while the dot-dash (red) and dotted (blue) lines show the impact of increasing the stellar magnetic field by factors of 10 and 100, respectively. Even in the solar case, $10^{26} \rm~erg~s^{-1}$ is available at $a = 0.1\rm~AU$, while in the most active case, energy deposition rates as high as $10^{30} \rm~erg~s^{-1}$ are possible  at $a = 0.05\rm~AU$.}
\end{figure}

In cases where the stellar magnetic field is substantially greater than the solar case, results are essentially independent of mass loss rate, for any mass loss rate which is reasonably solar-like. This is because, for such stars, magnetic effects are dominant in determining both $R_M$ and wind power available to the planetary magnetosphere for orbits inside $0.08 \rm~AU$.

\section{Discussion} 

\subsection{Observational Testing}

The magnetic heating model proposed above has observable ramifications, which can be tested using the relatively large and self-consistent data set available from the {\it Kepler} mission. In particular, as seen in Figure~2, the magnetospheric heating rate in this model is strongly dependent on stellar magnetic field strength, encapsulated at photospheric levels in the parameter $B_* f_*$. In turn, $B_* f_*$ is closely related to stellar activity levels (Pizzolato et al. 2003), for which typical observational proxies are normalized soft X-ray luminosity or emission in the cores of the Ca II H and K lines (Hall 2008). Although these particular proxies are generally unavailable for most stars with hot Jupiters (only 11 of the hot Jupiter host stars have extant Ca II measurements), I will follow Basri et al. (2011) and others in adopting the {\it Kepler} Combined Differential Photometry Precision (CDPP) as a measure of stellar photometric variability, and hence activity. A first order attempt was made to remove photometric noise from the CDPP by subtracting (in an $rms$ sense) the simple model CDPP derived in Machalek {\it et al.} (2011, see also Gilliland {\it et al.} 2011). Corrections were minor in most cases, and do not substantially affect my conclusions.

Approximately following Demory \& Seager, I used the {\it Kepler} Planet Candidate Data Explorer (http://planetquest.jpl.nasa.gov/kepler) to select all KOI planet candidates with $8R_\earth < R_{\rm PL} < 22R_\earth$, discarding those with no reported value for CDPP, resulting in 153 candidates. Figure 3 shows the distribution of CDPP as a function of planetary radius, with the constant radius level of $0.87 R_J$ seen by Demory \& Seager (2011) indicated by the red vertical line. While large planetary radii are associated with both active and inactive stars, small planetary radii are apparently only associated with inactive stars. I quantified this observation by applying a Kolmogorov-Smirnov test (Hollander \& Wolfe 1973) to test the hypothesis that the two groups represent the same underlying population. The K-S test indicates a probability $p = 0.03$ that the hypothesis is true, so we can conclude that a correlation exists between inflated radii and stellar activity. However, there is clearly considerable scatter in such a correlation (see Figure 3), as one might expect considering the impact of other parameters such as planetary composition, the strength and morphology of the planetary magnetic field, noncircular orbits, varying orbital alignments, and possible variations in the efficiency of wind energy deposition into the planetary interior. 

\begin{figure}
\epsscale{1.0}
\plotone{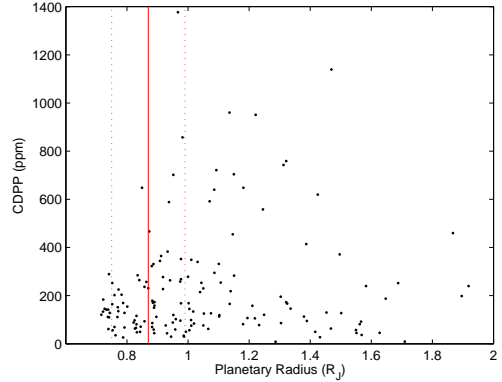}
\caption{Corrected CDPP values as a function of planetary radius for hot Jupiters ($8 R_\earth < R_{PL} < 22 R_\earth$) selected from KOI. CDPP values, corrected for non-astronomical noise sources, serve as a proxy for stellar activity. The vertical red line corresponds to the constant radius of $0.87 \pm 0.12 R_J$ found by Demory \& Seager (2011) for low incident stellar flux, and the dotted lines show the $\pm 1 \sigma$ levels. Visually, the populations vary at about that point, in the sense that while large planetary radii are associated with both active and inactive stars, small planetary radii are only associated with inactive ones.}
\end{figure}

\subsection{Possible Mechanisms}

We expect that, as in our solar system, exoplanet magnetospheres will be electrically connected to the upper ionosphere via a system of field aligned currents (FAC), the net result of which is to map the open solar wind magnetic field lines to the polar ionosphere and produce an electric potential structure across the ionosphere (Lysak 1980, Zhang {\it et al.} 2011). The scale of the potential difference can be estimated from $\Delta V \sim 2 \eta v_w B_w R_M$, where $\eta$ is a scale factor of order unity. In the solar system, taking $\eta = 1$ gives $\Delta V_\earth \sim 180 \rm~kV$ and $\Delta V_J \sim 2.2 \rm~MV$, both of which are of the correct order of magnitude compared to observations (Singh {\it et al.} 2007, Cowley {\it et al.} 2003). For a hot Jupiter located at $a = 0.08 \rm~AU$, the resulting potential difference is $\Delta V = 320 \rm~MV$. 

From the ionosphere, energy can be transferred downwards by a number of mechanisms (Vasyliunas, 2010), but here I will consider only two. First I consider auroral particle precipitation, in which relativistic particles are accelerated downwards from the ionosphere. While the details of the acceleration mechanism are complex and somewhat obscure (Olsson \& Janhunen, 2003), an order-of-magnitude energy estimate can be obtained using the potential difference estimated above. However, even the extremely relativistic $100 \rm~MeV$ electrons which might be expected for a hot Jupiter will only penetrate down to about the 50 millibar level (Berger {\it et al.}, 2005), which is clearly insufficient. Some of this energy will conduct to lower parts of the atmosphere and interior, but it is unclear how efficient such processes might be, and thus energetic particle precipitation is unlikely to contribute substantially to interior heating although it will increase conductivity in the polar regions of the atmosphere. 

The second mechanism for energy transport I will consider involves the planetary global electric circuit (GEC). On Earth, the ground and ionosphere together form what is essentially a leaky capacitor with the atmosphere acting as a dielectric. The voltage difference of $\sim 240 \rm~kV$ between the ground and ionosphere is maintained by a combination of thunderstorm action ($\approx 80\%$) and the solar wind-driven potential ($\approx 20\%$), and the circuit is closed through the fair-weather load resistance of the atmosphere, which is approximately $250 \Omega$ (Tinsley, Burns, \& Zhou, 2007). The total power involved in the terrestrial GEC is thus of order $2.3 \times 10^{15} \rm~erg~s^{-1}$, and can be an order of magnitude larger during substorms and other periods of enhanced solar activity. The existence of a GEC on giant planets in the solar system is debated (for contrasting views, see Aplin 2006 and Simoes {\it et al.} 2012), primarily because the low conductivity ($\sigma < 10^{-14} \rm~S~m^{-1}$, Whitten {\it et al.} 2008) in the upper Jovian and Kronian atmospheres renders it problematic; however, the higher temperatures anticipated in hot Jupiter atmospheres removes this obstacle.

Although lightning may be available to generate potential differences on hot Jupiters, as it is on Earth (Helling, Jardine, \& Diver 2012), for simplicity's sake I consider only the FACs, which as noted above can be expected to produce a cross polar cap potential of as much as $\Delta V = 320 \rm~MV$ between the day and night hemispheres on the canonical hot Jupiter with a semimajor axis $a = 0.08 \rm~AU$. This circuit can potentially  be closed either via the ionosphere or the planetary atmosphere and highly conductive interior. However, the conductivity of the ionospheric route is dominated by the Pedersen conductivity of the ionospheric layer, which for reasonable estimate of ionospheric parameters ($n_e \sim 10^4 - 10^5 \rm~cm^{-3}$, $T \sim 10^3 - 10^4 \rm~K$) is roughly five orders of magnitude lower than the conductivity in the radial direction. Accordingly, the resistance of the radial GEC circuit can be estimated using the calculated conductivity profile of Batygin, Stevenson, \& Bodenheimer (2011, $\epsilon = 0$ model as corrected on {\it astro-ph}) to be $\approx 1.7 \mu \Omega$. If we assume the canonical conversion efficiency of 1\% seen in solar system magnetospheres (Vasyliunas \& Song 2005), such a circuit leads to Joule heating levels of as much as $6 \times 10^{27} \rm~erg~s^{-1}$, more than sufficient to sink the available magnetospheric power. Figure~4 shows the resulting heating profile for an evolved $1 M_J$, $T_e = 1400 \rm ~K$ planet in a form which can be directly compared to Figure~4 of Batygin, Stevenson, \& Bodenheimer (2011); the profile from the GEC model is comparable to that of the Ohmic heating model at all pressure levels. Figure~4 also shows the boundary between the radiative and convective portions of the planetary interior, from which heating levels in the latter can be readily calculated. For the model shown, heating below the $1 \rm~kbar$ level is $2.1 \times 10^{25} \rm~erg~s^{-1}$, or $0.06 \%$ of the intercepted stellar power $L_p$. This is entirely compatible with the required heating levels calculated by Burrows {\it et al.} (2007).

Note that this model presumes a planetary $T_{e} = 1400 \rm K$; the effect of lowering this temperature to $1000 \rm K$ is to raise the resistance by two orders of magnitude, which would lower the Joule heating rate by a similar amount, and which mimics the situation expected on solar system gas giants. The GEC heating mechanism is thus sensitive to the temperature structure of the atmosphere, but will generally be more efficient as semimajor axis $a$ decreases (and thus atmospheric temperature and conductivity increase). One criticism of the proposed model is that the presence of an electrically insulating layer (as posited by Ohmic heating models) would substantially reduce or eliminate GEC heating of the interior. Such as layer might arise through the presence of an atmospheric inversion layer produced by opacity effects (Hubeny, Burrows, \& Sudarsky 2003) or a thermoresistive instability (Menou 2012), and evidence for it at the $10 - 100 \rm~mbar$ level has been detected in some, though not all, hot Jupiters (Knutson, Howard, \& Isaacson 2010). However, it is likely that exoplanet atmosphere conductivity profiles are spatially inhomogeneous; in particular, as noted above, nonthermal ionization due to auroral precipitation will tend to increase conductivity in the polar regions that are most critical to the proposed model.

\begin{figure}
\epsscale{1.0}
\plotone{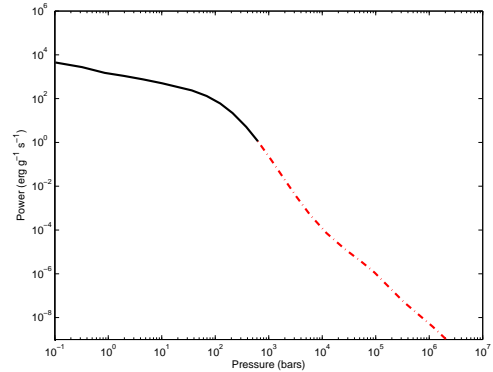}
\caption{A GEC heating model showing energy dissipation per unit mass, calculated for the conductivity profile corresponding to the evolved $1 \rm~M_J$ planetary model shown in Batygin {\it et al.} (2011). The black (solid) line corresponds to the radiative part of the planetary interior, while the red (dot-dash) line represents the convective inner region. Total heat input below the $1 \rm~kbar$ level is $2.1 \times 10^{25} \rm~erg~s^{-1}$.}
\end{figure}




\acknowledgments

I am grateful to the Whitaker Foundation for their support of the Whitaker Chair at Florida Gulf Coast University. This work has made use of the Exoplanet Database at exoplanets.org, the {\it Kepler} Planet Candidate Data Explorer at planetquest.jpl.nasa.gov, and NASA's Astrophysics Data System. Funding for the {\it Kepler} mission is provided by the NASA Science Mission Directorate, and this work was in part funded by the {\it Kepler} Participating Science Program grant NNH09CE70C.



{\it Facilities:} \facility{{\it Kepler}}




\clearpage



\end{document}